\documentclass[twocolumn,floatfix,aps,amsmath,amssymb,amsfonts,10pt,showpacs,prd,nofootinbib]{revtex4-1}

\usepackage[T1]{fontenc}		
\usepackage[utf8]{inputenc}		
\usepackage{epsfig}
\usepackage{hyperref}
\usepackage{listings}
\begin{document}

\title{A QAOA approach with fake devices: A case study \\
	    for the maximum cut in ring graphs}

\author{Wilson R. M. Rabelo}
\email{rabelo@ufpa.br}
\affiliation{Instituto de Tecnologia, Universidade Federal do Pará, Belém, PA, Brasil}
\author{ Sandra D. Prado}
\author{Leonardo G. Brunnet}
\affiliation{Instituto de Física, Universidade Federal do Rio Grande do Sul, Porto Alegre, RS, Brasil}


\begin{abstract}
	The quantum approximate optimization algorithm (QAOA) can require considerable processing time for developers to test and debug their codes on expensive quantum devices. One avenue to circumvent this difficulty is to use the error maps of quantum devices, where a local simulator can be automatically configured to mimic an actual device backend.
	In our work, we evaluated some error maps of quantum devices, known as fake devices, that are freely available in the cloud. The QAOA and the problem of maximum cut in 2-regular connected graphs, known as ring of disagrees, were used as tools for the noise analysis. The approximation ratio, the expectation energy and the probability of success for this problem have been evaluated in two scenarios. First, the quantities were studied through noisy simulations using fake devices. Second, error mitigation methods such as optimization levels and translation (connectivity mapping) of the original problem were applied. These results were then compared with the analytical solution of the ring graph. The study shows that error mitigation methods were crucial in obtaining better results for the expectation value of the energy, the approximation ratio, and the probability of success for the ring graphs.
\\	
\\
{\bf Keywords:} {max-cut; QAOA; error maps; noisy simulations; quantum algorithms.}
\end{abstract}

\maketitle

\section{Introduction}
\label{intro}
In recent years, we have observed an explosion of quantum device prototypes available to academic researchers or commercial laboratories. Current quantum computers have already demonstrated the ability to outperform classical computers in certain mathematically designed tasks \cite{Arute, Zhu}. However, our understanding of how to develop noise-resilient algorithms that can run on near-term quantum resources remains limited \cite{NISQ}.

The quantum approximate optimization algorithm (QAOA), which follows a hybrid classical-quantum scheme, has been designed to solve combinatorial optimization problems on a gate-based quantum computer \cite{Farhi-2014, Peruzzo-2014}. Combinatorial optimization is the search for the object that minimizes a cost function among a finite number of objects. The QAOA is based on a reformulation of combinatorial optimization in terms of finding an approximation to the ground state of a Hamiltonian by adopting a specific variational ansatz for the trial quantum state. This ansatz is specified in terms of a gate set and involves $2p$ variational parameters that must be optimized by running a minimization algorithm on a conventional computer. One such combinatorial optimization problem, known as the maximum cut (max-cut) problem \cite{max-cut}, attracted much attention early on in the quantum computing community. 
The max-cut problem is usually mathematically represented as the problem of partitioning the vertices of an undirected graph into two sets such that the number of vertices connecting both sets is maximized. This problem is known to be NP-complete \cite{NP-maxcut}. 
There has also been a significant amount of follow-up work on the QAOA algorithm for the max-cut problem\cite{Farhi-QAOA-Supremacy,Review-2022-QAOA}. 

In this work, we contribute three aspects to analyzing noise in local simulations with the QAOA approach, dealing with maximum cut problems for ring graphs. In simulations, we use fake devices, that is, error maps of real quantum devices that are available free of charge in the cloud. First, we performed noise mapping in simulations using AerSimulator (Qiskit) to obtain the maximum cut in a ring graph, with noise maps from three quantum devices of different sizes: Fake-Lagos, Fake-Kolkata, and Fake-Washington, with $7, 27, 127$-qubits, respectively. Second, we use error mitigation methods to analyze the performance of these devices for maximum cut problems in ring graphs. In addition, we analyze a particular question when we have a ring graph with a symmetry that can be mirrored in the topology of the quantum device.
Finally, we provide an approach for analyzing a specific optimization problem by carrying out only local simulations using fake devices available in the cloud. Thus, we circumvent the need for access to relatively expensive real quantum devices in the first analysis. This can be important for developers when creating, testing, and debugging their codes without needing real quantum devices.

The paper is organized as follows: In Sect. 2, we introduce the maximum cut problem and the approximation ratio for algorithms and review the fundamental issues of QAOA. In Sect. 3, we discuss the ring graph model for the QAOA approach for $p \geqslant 1$. In Sect. 4, we discuss the noise models used and the error mitigation methods using the Qiskit framework. In Sect. 5, we present the results obtained using quantum simulator, the AerSimulator (Qiskit Aer). In Sect. 6 contains our conclusions from the work.

\section[maxcut]{Theoretical support}

\subsection{Approximation ratio and maximum cut}
Given a combinatorial optimization problem defined over n-bit binary strings in the form $\textbf{x}=x_1,\ldots, x_n$, where the objective is to maximize a given cost function $C(\textbf{x}): \{0,1\}^n \rightarrow \textbf{R}$, an approximative algorithm aims to discover a string $\textbf{x\textquotesingle}$ such that $C(\textbf{x\textquotesingle})$ falls within the factor $r^* (\leqslant 1)$ of the optimal solution. In the context of maximization, the algorithm aspires to ascertain a string $\textbf{x\textquotesingle}$ that attains a desired approximation ratio $r^*$, as defined by the equation:
\begin{equation}\label{ratio_r}
	\frac{C(\textbf{x\textquotesingle})}{C_{\max}} \geqslant r^*,
\end{equation}
with $C_{max} = {\max_{x}} C(x)$. Ideally, the value of this approximation ratio should approximate unity \cite{Cormen}.

Maximum cut – Our focus in this article is on analyzing the maximum cut (max-cut) in graphs, a combinatorial optimization problem. Given a graph $G = (V,E)$ with a vertex set $V$ and edge set $E$, and $w_{ij}$ denoting the corresponding weight for edge $(i,j)\in E$, which links vertices $v_i$ and $v_j$, a cut is a partition of the elements of $V$ into two disjoint subsets $S$ and $\bar{S}$, where $S \cup \bar{S} = V$. The max-cut problem consists in finding a cut $V$ that maximizes $\sum_{(i,j)\in E} w_{ij}$, with $w_{ij} > 0$, $w_{ij}=w_{ji}$.
For every vertex $v_i \in V$, let us associate a variable $x_i \in \{-1,+1\}$.
Given an arbitrary cut $V$, let us define $x_i = 1$ if $v_i \in S$ and $-1$ otherwise. Then, the max-cut problem is equivalent to the following quadratic binary
program,
\begin{equation}\label{maxcut}
	\max: ~ \sum_{(i,j)\in E} \frac{w_{ij}}{2}(1-x_ix_j) ,   
\end{equation}
subject to $x_i \in \{-1,+1\}$, $\forall v_i \in V$, $\forall (i,j) \in E$. The term "weighted max-cut" is employed when the edges possess arbitrary weights $w_{ij}$. A special case arises when the weighted max-cut entails $w_{ij}=1$ for any $(i,j) \in E$.

\subsection{The QAOA}

QAOA was first introduced by Farhi et al. \cite{Farhi-2014} and was designed to determine approximate solutions to combinatorial optimization problems. In Ref.~\cite{Farhi-2014}, the max-cut was the first example studied using the QAOA approach. In the following, we will summarize QAOA in the following points:

(1) (Cost function and the Hamiltonian ${\hat H}_C$): Let there be a combinatorial optimization problem defined by an objective function represented as $C(\textbf{x})$, such as the max-cut in equation (\ref{maxcut}). We shall define a cost Hamiltonian ${\hat H}_C$, such that its lowest energy states encode the solution to the optimization problem. For the max-cut, the Hamiltonian ${\hat H}_C$ is diagonal in the computational basis, where each binary variable $x_i \in \{+1,-1\}$ of the max-cut is mapped to the Pauli $Z_i$-operator. Let us establish a mixing Hamiltonian ${\hat H}_M$ that does not commute with ${\hat H}_C$ which helps guide the optimization in Hilbert space towards the ground state of $H_C$. Therefore:
\begin{align}
	\label{H_C}	
	{\hat H}_C & = \frac{1}{2} \sum_{(i,j)\in E} w_{ij}(I-Z_iZ_j), \\
	\label{H_M}
	{\hat H}_M & = \sum_{j \in V} X_j,
\end{align}
where $I$ represents the identity operator, and $Z_j$ ($X_j$) is the Pauli $Z$ ($X$)-operator acting on qubit $j$-th. It is noteworthy that the cost function of the max-cut corresponds to a classical Ising Hamiltonian, and the edge (i, j) contributes a value of $w_{ij}$ if and only if the spins $Z_iZ_j$ are oriented antiparallel. Beyond the max-cut, several NP-hard combinatorial problems can be mapped to an Ising-type Hamiltonian. For further details, see Refs.~\cite{QUBO,IsingFormulations}.

(2) (Preparation of the initial state): The circuit start with the initialization of $n$ qubits in the standard state $|0\rangle^{\otimes n}$, with $n=|V|$ vertices and $m=|E|$ edges. Subsequently, we apply $n$ Hadamard gates to each of the $n$ qubits. Consequently, we obtain an initial quantum state that represents a balanced superposition of all $2^n$ possible problem strings to be analyzed. Thus,
\begin{equation}\label{initial-state}
	|\psi_{in}\rangle = H^{\otimes n }|0\rangle=|+\rangle^{\otimes n}= \frac{1}{\sqrt{2^n}} \sum_{\textbf{x}\in\{0,1\}} |\textbf{x}\rangle,
\end{equation}
where $n$ denotes the quantity of qubits, with $n=|V|$ vertices. It is worth noting that this initial quantum state also corresponds to the highest energy state of the mixing Hamiltonian ${\hat H}_M.$

(3) (Circuit ansatz): The next step is to construct the circuit ansatz, which is defined by the following unitary transformations:
\begin{align}
	\label{unitary-ansatz1}
	{\hat U}_C(\gamma) & = e^{-i \gamma {\hat H_C}} = \prod_{i=1,j<i}^{n} R_{Z_iZ_j}(-2w_{ij}\gamma), \\
	\label{unitary-ansatz2}
	{\hat U}_M(\beta)  & = e^{-i \beta {\hat H_M}} = \prod_{i=1}^{n} R_{X_i}(2\beta),
\end{align}
\begin{figure}[t]
	\centering
	\includegraphics[scale=0.6]{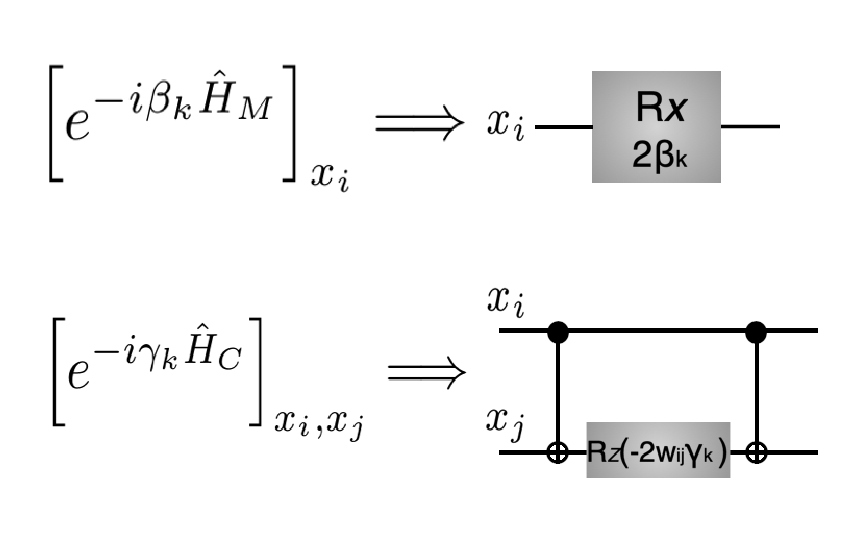}
	\caption{The mixing layer, denoted ${\hat U}_M(\beta_k)$ (above), and the cost layer, denoted ${\hat U}_C(\gamma_k)$ (below), along with their representations in terms of logical gates and the corresponding variational parameters, $\beta_k$ and $\gamma_k$.}
	\label{uni-qaoa}
\end{figure}
where $\gamma$ and $\beta$ are denoted as the variational parameters of the circuit. In the literature, ${\hat U}_C(\gamma)$ and ${\hat U}_M(\beta)$ are recognized as the cost and mixing layers, respectively. A single layer of the QAOA consists of two layers: the cost and the mixing, which can be combined into additional layers to increase the depth $d$ of the quantum circuit. In Figure \ref{uni-qaoa} we illustrate how each layer can be implemented in terms of logical gates. The mixing layer ${\hat U}_M(\beta)$ can be realized by a sequence of 1-qubit rotation gates $R_x(2\beta_k)$ \cite{NC10}, while for the cost layer ${\hat U}_C(\gamma)$ we use a pair of CNOT gates \cite{NC10} and a 1-qubit $R_z(-2w_{ij}\gamma_k)$ rotation gate \cite{NC10} to implement the two-qubits interaction for the cost layer.

(4) (Output state and optimal parameter set) QAOA$_p$ is parameterized within a set of $2p$ variational parameters, where the index $p$ refers to the layer level of the QAOA circuit with a total number of layers $p\geqslant 1$. These parameters are denoted as $\gamma = (\gamma_1,\ldots,\gamma_p)$ and $\beta = (\beta_1,\ldots,\beta_p)$, such that $\gamma_k$ is in the interval $[0,2\pi]$ and $\beta_k$ is in the range $[0,\pi]$, for $k=1,\ldots,p$. The output quantum state (trial state) of the circuit is defined as follows:
\begin{equation}
	\left|\psi_{p}(\gamma,\beta)\right\rangle=e^{-i\beta_{p}\hat{H}_{M}}e^{-i\gamma_{p}\hat{H}_{C}}\cdot\cdot\cdot e^{-i\beta_{1}\hat{H}_{M}}e^{-i\gamma_{1}\hat{H}_{C}}\left|\psi_{in}\right\rangle .
\end{equation}
Note that the $p$ parameter determines the number of parameters independent of the trial state. The expectation value of the Hamiltonian $H_C$ with respect to the ansatz state $\left|\psi_{p}(\gamma,\beta)\right\rangle$ is evaluated by repeated measurements of the final state in the computational basis, then
\begin{equation}
	F_p(\gamma,\beta) = \langle\psi_{p}(\gamma,\beta)| {\hat{H}_C} |\psi_{p}(\gamma,\beta)\rangle .
\end{equation}

A classical optimization algorithm (conventional digital computer) is used to iteratively update the parameters $\gamma$ and $\beta$. The goal of this classical procedure is to find the optimal set of parameters, denoted as $(\gamma^*,\beta^*)$, such that the expectation value of $F_p(\gamma,\beta)$ is maximized: 
\begin{equation}
	F^*=F(\gamma^*,\beta^*) = \arg\max_{\gamma,\beta} F_p(\gamma,\beta) .
\end{equation}
At the end of the optimization, the approximation ratio $r$ is given by
\begin{equation} 
	r = \frac{F^*}{C_{\max}} ~,
	\label{F-ratio}
\end{equation}
Optimal state $|\psi_{p}(\gamma^*,\beta^*)\rangle$ will encode the solutions to the optimization problem, i.e., once $\gamma^*$ and $\beta^*$  have been determined, repeated measurement in the computational basis of the state $|\psi_{p}(\gamma^*,\beta^*)\rangle$ of the quantum computer yields a sample of $x$'s, with probability $P(x)=|\langle x|\gamma^*,\beta^*\rangle|^2$.

\begin{figure}[h]
	\centering
	\includegraphics[scale=0.34]{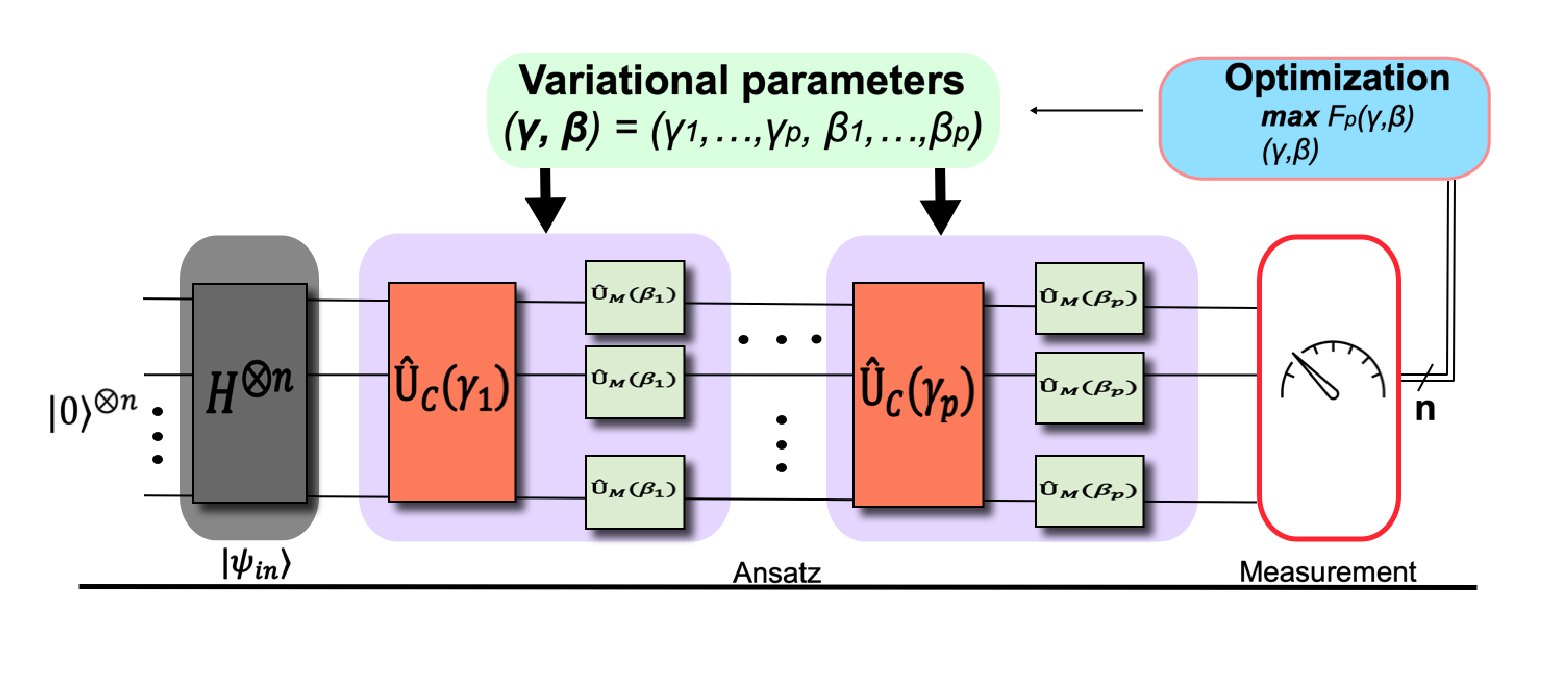}
	\caption{QAOA workflow diagram for $p$ layers.}
	\label{Diagram-QAOA}
\end{figure}

In summary, the usefulness of the ansatz circuit is to drive the system to a quantum state such that if we perform a series of measurements on the computational basis of it, we will with high probability obtain a classical bit string that is approximately $r^*$ from the optimum.
A schematic of all QAOA algorithm steps is shown in Figure \ref{Diagram-QAOA}. 

\section{The ring model}
\label{sec:example}
Before we get into the numerical tests and noise analysis, let us take a quick look at the QAOA approach applied to max-cut on a toy example for $p=1$ by analyzing it analytically. Let us take a 2-regular connected graph, i.e. the ring of disagrees \cite{Farhi-2014}. For simplicity, we assume $n=m=4$, $w_{ij}=w_{ji}=1$ and $x_i \in \{-1,+1\}$. Note that the maximum cut in this particular case is $n$ or $n-1$ if $n$ is even or odd, respectively, and only one edge in $\langle i,j \rangle$ of the equation \ref{H_C} needs to be evaluated. For ring graphs, we only need to evaluate an expectation value $\langle H_{i,j}\rangle$ (subgraph) for one edge instead of all of them. This follows from the fact that for the ring model there is only one type of segment subgraph of $2p+2$ connected vertices, as shown in Refs.~\cite{Farhi-2014,Wang-2018}. It is important to emphasize the simplification when there are multiple isomorphic subgraphs, which allows for a speedup of the algorithm.
Then, using Equation \ref{maxcut},
\begin{align}
	\label{H_C-n-4}	
	C(x) &= \sum_{(i,j)\in E} C_{i,j} ~, 	
\end{align}
with $C_{ij}=\frac{1}{2} (1-x_ix_j)$.
Following the steps of the QAOA$_1$ workflow in Figure \ref{Diagram-QAOA}, with $p=1$, and analyzing only one edge $C_{ij}$, we have:

(1) The objective function $C_{ij}$, referring to the edge $\langle i,j\rangle$ for a Hamiltonian cost $H_{ij}$. From Equations \ref{H_C} and \ref{H_M}, the Hamiltonian cost and mixing are given by:
\begin{align}
	{\hat H_{ij}}  &= \frac{1}{2} (I-Z_iZ_j)~, \\
	{\hat H}_M^{ij}     & = \sum_{i} X_i,
	\label{Hc}
\end{align}

\begin{figure}[h]
	\centering
	\includegraphics[scale=0.45]{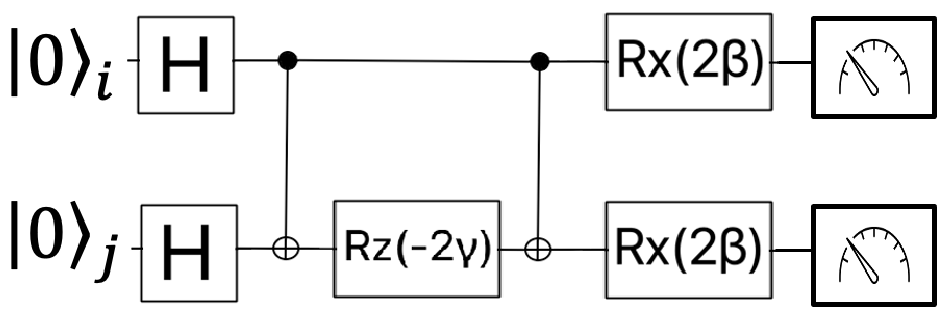}
	\caption{Ansatz circuit for the edge $\langle i,j \rangle$ for the 2-regular graph, connected with n=4.}
	\label{n-4-ring}
\end{figure}

(2) The initial state after applying the two Hadamard gates is: $|\psi_{in}\rangle = |+\rangle_i \otimes |+\rangle_j$, for a single layer of QAOA$_1$, see Figure \ref{n-4-ring}.

(3) The ansatz circuit for this case is given by the Equations \ref{unitary-ansatz1} and \ref{unitary-ansatz2}:
\begin{align}
	{\hat U}_C(\gamma) & = e^{-i \gamma {\hat H}_C} = R_{Z_iZ_j}(-2\gamma), \\
	{\hat U}_M(\beta)  & = e^{-i \beta  {\hat H}_M} = \prod_{i=1}^{n=2} R_{X_i}(2\beta).
\end{align}
Note that the constant part of the Hamiltonian ${\hat H}_C$, which is $e^{-i \gamma I/2}$, places only a global phase on the system state and thus can be suppressed. Note also that the unitary transformations ${\hat U}_C(\gamma)$ and ${\hat U}_M(\beta)$ apply a phase and a mixing to all probability amplitudes of the quantum state of the system.

(4) The expectation value of the Hamiltonian ${\hat H_{ij}}$ with respect to the ansatz state
$\langle\psi(\gamma,\beta)|{\hat H_{ij}}|\psi(\gamma,\beta)\rangle$ is given by:
\begin{align}
	\label{n-2-Fp-f}
	F^{ij} & = \frac{1}{2} \left[ 1 +  \sin(4\beta)\sin(\gamma)\cos(\gamma) \right].
\end{align}
\begin{figure}[h]
	\centering
	\includegraphics[height=60mm, width=80mm]{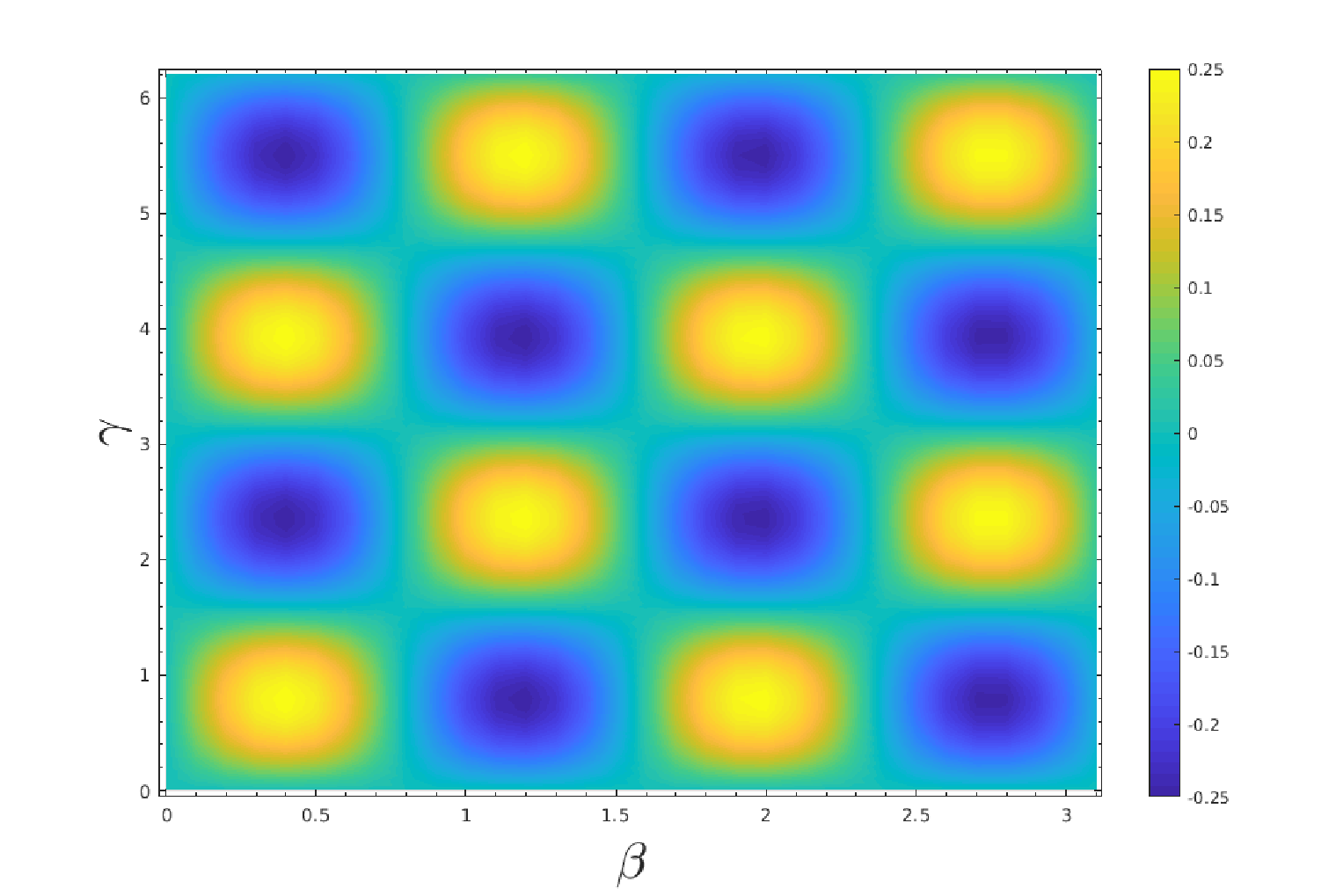}
	\caption{Contour plot of expected energy for the ring model QAOA states when $p=1$.}
	\label{plot-n2-QAOA}
\end{figure}
Therefore, the expectation value of ${\hat H}_C$ in the QAOA is:
\begin{align}
	\label{n-2-Fp-f}
	F(\gamma,\beta) & = m F^{ij}. 
\end{align}
Note that the optimal parameters for the above example are given by $(\gamma^* = \pi/4,\beta^*=\pi/8)$. From the equation \ref{F-ratio}, the approximation ratio is given by $r^* = {F^*}/{C_{\max}}=3/4$. Figure \ref{plot-n2-QAOA} shows the contour plot for the 2-regular ring model for $n=4$. As we can see from the Refs.~\cite{Zhou-2020,Pendas-2022}, the energies or parameter pairs $(\gamma_k,\beta_k)$, are difficult to optimize due to the various minima and maxima of the system energies and the 0-value saddle points between them, which for $p>1$ can lead classical optimizers to get trapped in these locations.

Observe that we have the form of ${\hat H}_C$ and ${\hat H}_M$ bounding the angles $\gamma_k$ and $\beta_k$, where ${\hat H}_C$ is a diagonal matrix with integer eigenvalues, so $e^{-i\gamma_k {\hat H}_C}$ is $2\pi$-periodic in $\gamma_k$. For ${\hat H}_M$ we have $e^{-i(\pi+\beta_k)X}=e^{-i\pi X} \cdot e^{-i\beta_k X}=-I \cdot e^{-i\beta_k X}$, so $e^{-i\beta_k X}$ is (up to a global phase) $\pi$-periodic in $\beta_k$ and reduces the search space for the optimal angles. 

From Figure \ref{plot-n2-QAOA} we can infer that for the specific case of max-cut we can still limit the range of 
$\beta_k$ and $\gamma_k$, because ${\hat H}_C$ and ${\hat H}_M$ have a real valued and satisfy the time reversal symmetry. Therefore, $F(2\pi-\gamma,\frac{\pi}{2}-\beta)=F(\gamma,\beta)$, with the angles $\beta_k$ and $\gamma_k$ varying by $[0,\pi/2]$ and $[0,\pi]$, respectively, and using these symmetries we reduce the degeneracies of the optimal parameters.
In general, for the weighted max-cut case, the angle $\gamma_k$ can no longer be restricted to the interval $[0,2\pi]$.

The above findings are consistent with Refs.~\cite{Farhi-2014} and ~\cite{Wang-2018} for the max-cut case of a 2-regular, connected graph.  In this case, the ring of disagrees for even n, the optimal partition is to include every other vertex into one set and the rest into the other set, hence $C_{\max} = n$. 
It has been based on numerical evidence \cite{Farhi-2014, Wang-2018}, that the optimal expectation value is
\begin{align}
	\label{F*-ring}
	F^* & = -\frac{n(2p+1)}{2p+2}~, 
\end{align}
for all $p$. Thus the approximation ratio is
\begin{align}
	\label{r*-ring}	
	r^* & = \frac{F^*}{C_{\max}}=\frac{2p+1}{2p+2}~.
\end{align}
Note that, the approximation ratio on the problem of the ring of disagrees of even $n$ is independent of the problem size. 
For more details on the QAOA approach for the max-cut with $p=1$ on arbitrary graphs, see the Ref.~\cite{Wang-2018}.

\section{Noise model and error mitigation using Qiskit}

\subsection{Noise model simulations}
The quantum information science kit (Qiskit) is a Python software library for writing, editing, optimizing and running quantum circuits on quantum devices and local simulations \cite{Qiskit}. 
Qiskit Aer \cite{Qiskit-Aer}, a module of Qiskit, provides a simulator (AerSimulator) and tools for generating noise models to perform realistic noise simulations of the errors that can occur during execution on real devices. This allows developers to test and debug their code without the need for expensive quantum hardware.

\begin{figure}[h]
	\centering
	\includegraphics[height=60mm, width=80mm]{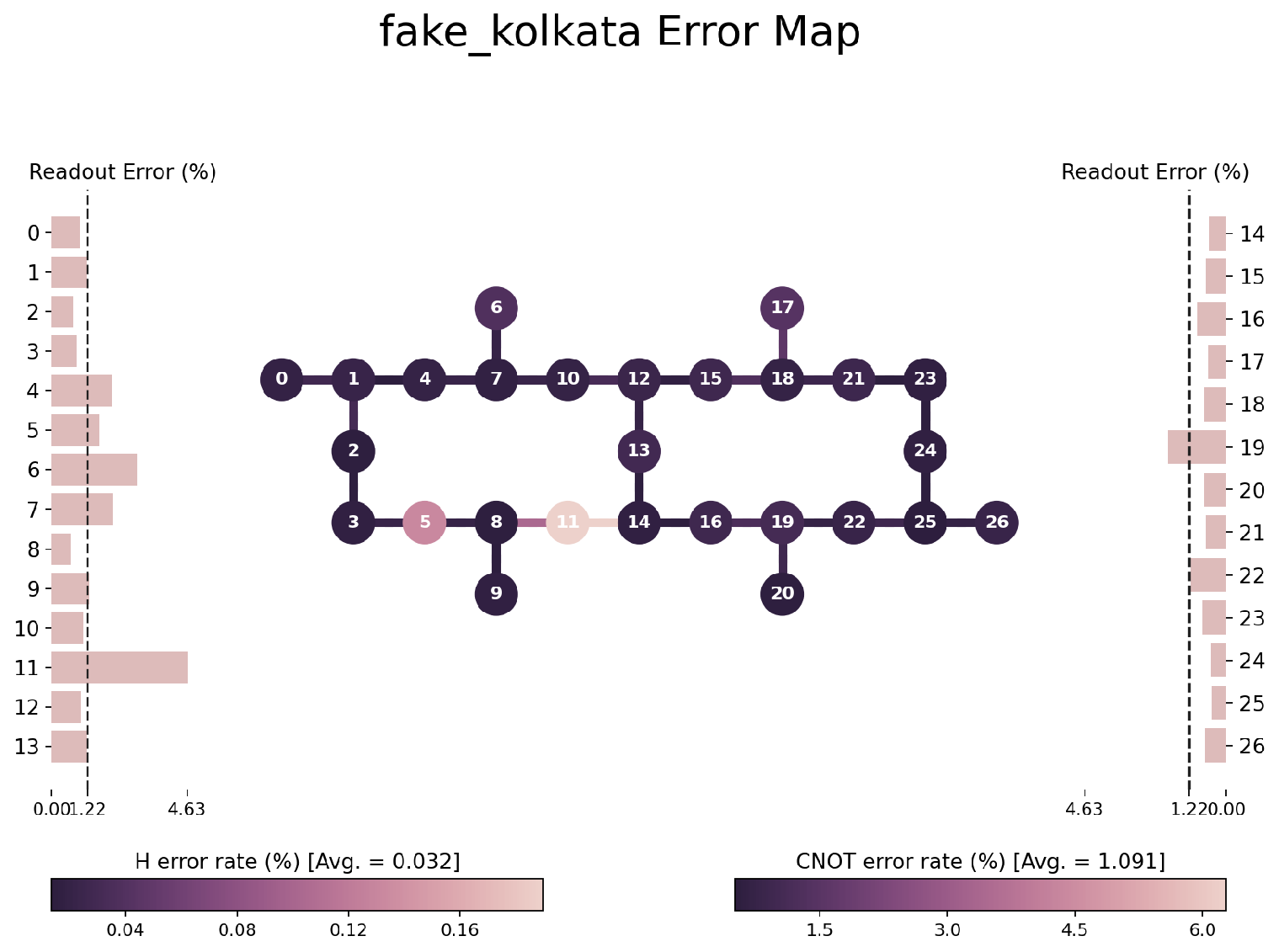}
	\caption{Error map for IBM's Fake-Kolkata, with 27-qubits \cite{IBMQ-Fake}.}
	\label{ibm-kolkata-error}
\end{figure}

\begin{figure}[h]
	\centering
	\includegraphics[height=60mm, width=80mm]{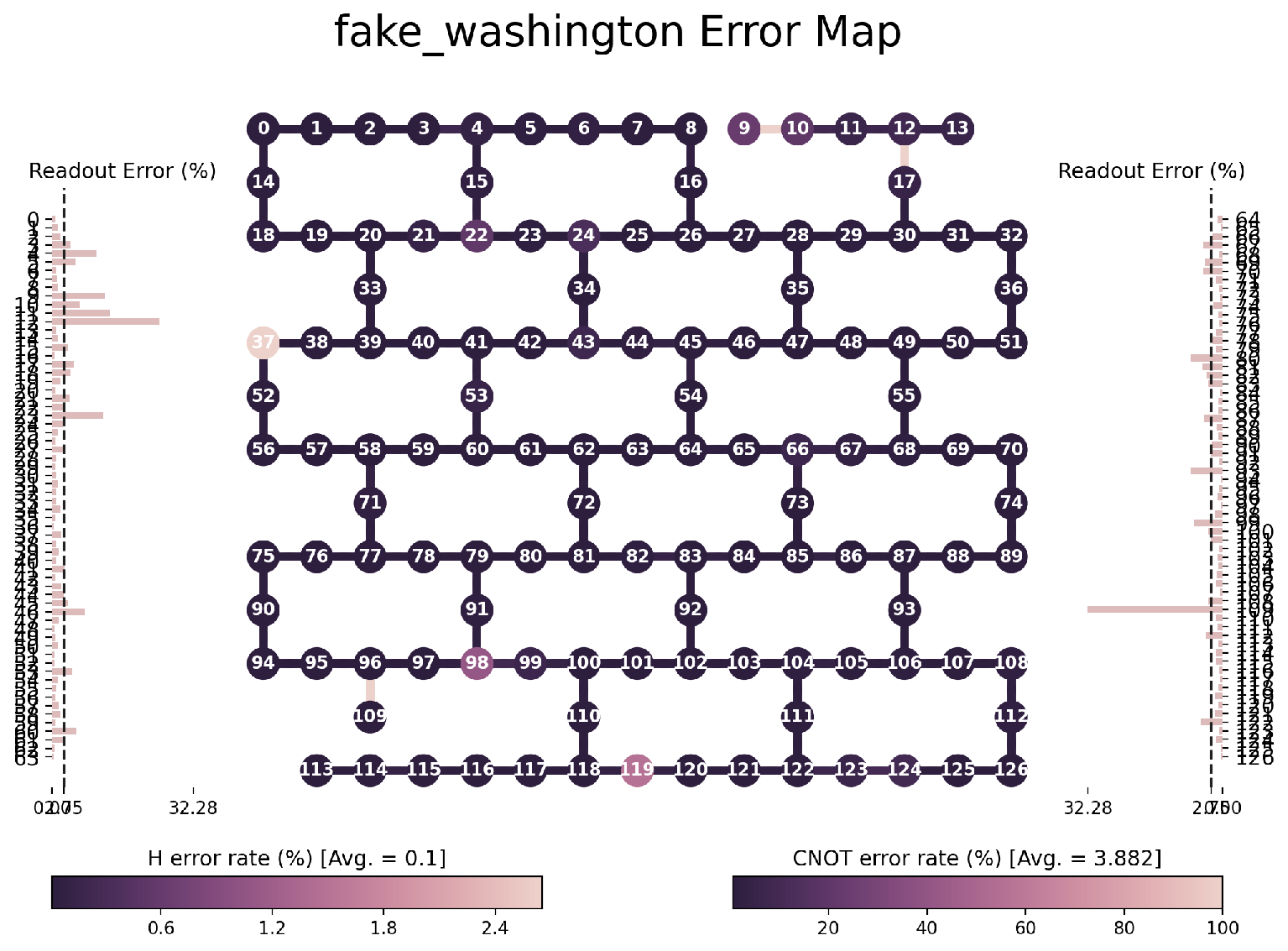}
	\caption{Error map for IBM's Fake-Washington, with 127-qubits \cite{IBMQ-Fake}.}
	\label{ibm-washington-error}
\end{figure}

It is noteworthy to emphasize that QAOA uses a classical computer to optimize the $2p$ parameters of a quantum circuit. This method can be computationally expensive given the execution time required for the detailed investigation of research problems addressed as $p$ grows. Therefore, one avenue to circumvent this difficulty is to use noise maps of quantum devices, which simulate these devices on conventional computers.
\begin{figure}[h]
	\centering
	\includegraphics[height=60mm, width=80mm]{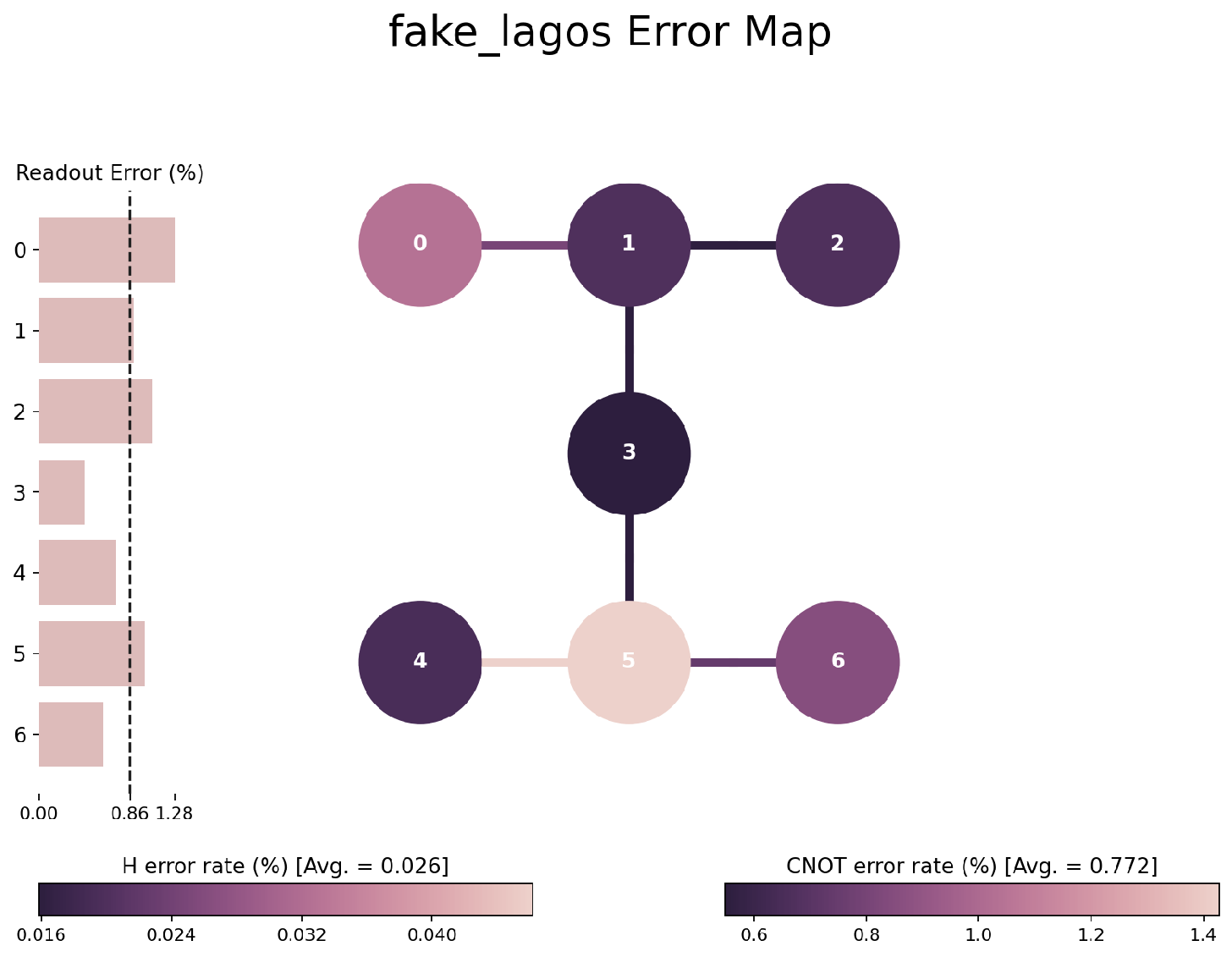}
	\caption{Error map for IBM's Fake-Lagos, with 7-qubits \cite{IBMQ-Fake}.}
	\label{ibm-lagos-error}
\end{figure}
The Qiskit Aer automatically furnishes simplified noise models of a real device. This model is devised using the calibration data reported from a device, and includes the following information: single-qubit gate errors\footnote{Single qubit depolarizing error followed by a single qubit thermal relaxation error.}, two-qubit gate errors\footnote{Two-qubit depolarizing error followed by single-qubit thermal relaxation errors on both qubits in the gate.} and single-qubit readout errors\footnote{Classical bit value obtained from measurements on individual qubits. For more details, see Ref.~\cite{Qiskit-DocEM}}.
According to the Qiskit documentation \cite{Qiskit-Aer}, the local simulator can be configured to mimic a real device (backend) using the \textit{from\_backend() method}.  The local simulator will use the basic device \textit{NoiseModel} for this backend, and the same basic gates and coupling map.
In this paper we consider error maps for the following devices: Fake-Lagos, Fake-Kolkata, and Fake-Washington, with $7, 27, 127$-qubits, respectively. The error rates on 1-qubit gates, 2-qubit gates and readout are reported in the error maps in Figures \ref{ibm-lagos-error}, \ref{ibm-kolkata-error}, and \ref{ibm-washington-error} \cite{IBMQ-Fake}.


\subsection{Error mitigation}
In this paper we will refer to error mitigation(EM) as a set of methods used to improve and modify a circuit during compilation in order to reduce errors.
It is noteworthy that this error treatment technique typically yields some classical preprocessing overhead in the overall runtime.
If the graph of a problem mirrors or is a subset of the quantum device architecture, then the problem can be seamlessly mapped onto the device. Otherwise, additional manipulations (swap gates) are required to map the problem to the chip's architecture. Therefore, circuit transpilation is an important task to find the minimum number of swap gates required to map a quantum circuit onto a given device. 
\begin{figure*}[t]
	\centering
	\includegraphics[width=0.95\textwidth]{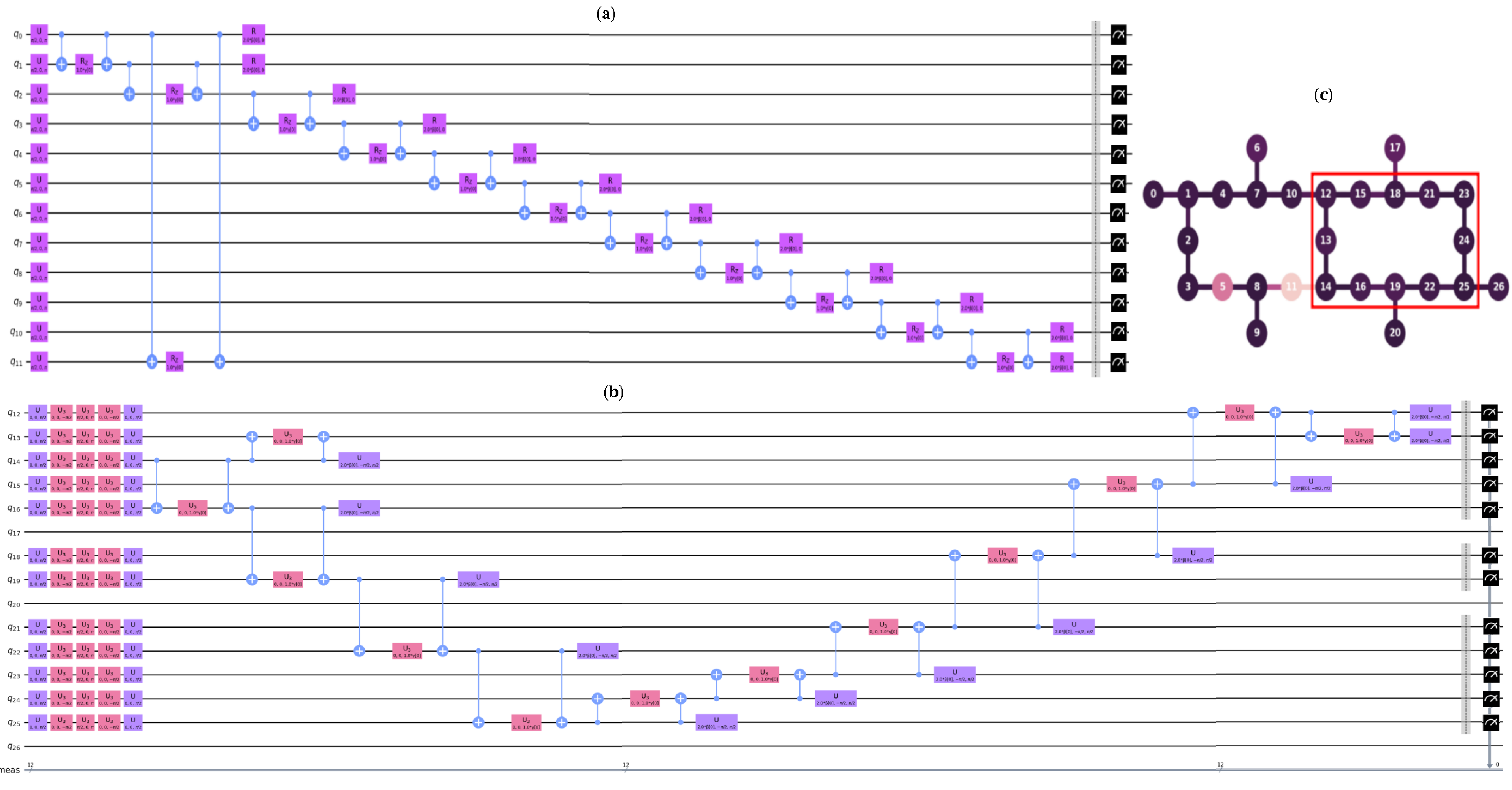}
	\caption{(a) For a ring graph with a size of $n = 12$, the virtual circuit (QAOA ansatz, p=1) has a depth $d = 36$ and a number of circuit operations of 60. (b) After the transpilation process with \textit{optimization\_level=3} for the device with the Fake-Kolkata error map, we have $d=40$ and a number of circuit operations of 108, with global phase ($2*\pi -6.0\gamma[0])$. (c) Indicated in red, the transpilation mirrors the ring graph onto the device. Due to the size of the figure, the symbols representing the measurements on each qubit have been arranged in a single column.}
	\label{Circuit-Virtual-n12}
\end{figure*}
The transpilation is the complex procedure of rephrasing a provided input circuit to match the topology of a particular quantum device and to refine the circuit for execution on noisy quantum systems. We work with the error maps of the Fake-Lagos, Fake-Kolkata, and Fake-Washington devices, performing simulations without and with error mitigation. Error mitigation was carried out using the following methods: \textit{optimization\_level} and \textit{translation\_method}.

In the following excerpts, code listings have been extracted from the Qiskit\footnote{Github: "Qiskit is an open-source SDK for working with quantum computers at the level of extended quantum circuits, operators, and primitives." \cite{Qiskit}} documentation and supplemented with the necessary integrations of EM methods.
\lstset{tabsize=2,basicstyle=\small, emph={optimization_level,translation_method},emphstyle=\underbar,frame=lines}
\begin{lstlisting}[caption={ Noise model without EM },label=code-EM-1]
	### Noise model without EM ###
	seed = 19999
	noisy_estimator = AerEstimator(
	backend_options={
		"method": "density_matrix",
		"coupling_map": coupling_map,
		"noise_model": noise_model,
		"basis_gates": basis_gates,},
	run_options={"seed": seed, "shots": 50000},
	transpile_options={"seed_transpiler": seed}
	,)
\end{lstlisting}
\begin{lstlisting}[caption={ Noise model with EM },label=code-EM-2]	
	### Noise model with EM ###
	seed = 17777
	noisy_estimator_EM = AerEstimator(
	backend_options={
		"method": "density_matrix",
		"coupling_map": coupling_map,
		"noise_model": noise_model,
		"basis_gates": basis_gates,},
	run_options={"seed": seed, "shots": 50000},
	transpile_options={"seed_transpiler": seed,
		"optimization_level": 3, 
		"translation_method": 'synthesis'},)		
\end{lstlisting}

\begin{figure}[h]
	\centering
	\includegraphics[height=60mm, width=80mm]{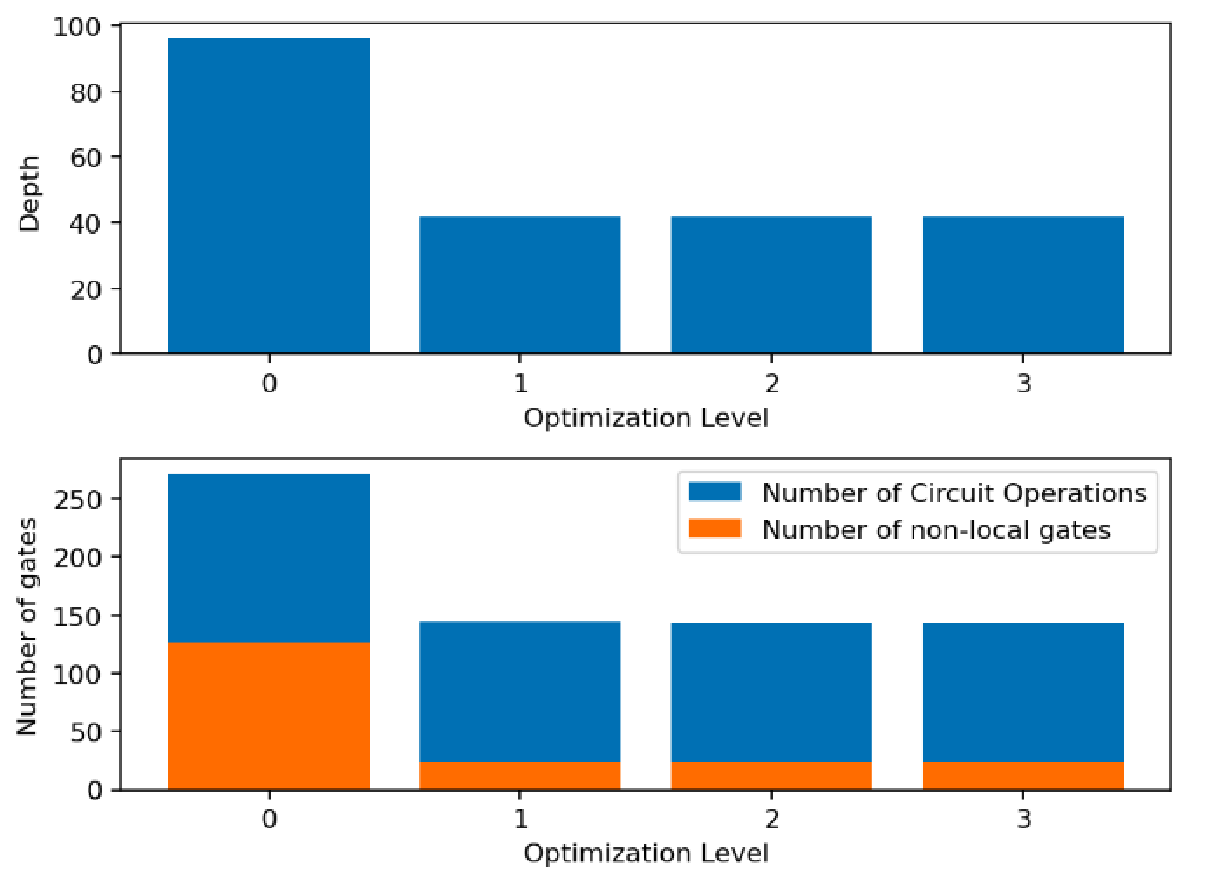}
	\caption{The distinct levels of optimization for depth, the number of circuit operations and the number of non-local gates, for $n=12$, with Fake-Kolkata device.}
	\label{Plot-n-12-Depth}
\end{figure}
To illustrate, consider a ring graph with a size of $n = 12$. The virtual circuit (QAOA ansatz, p=1) has a depth $d = 36$ and a number of circuit operations of 60, see the Figure \ref{Circuit-Virtual-n12}-(a).  For the transpilation process with \textit{optimization\_level=0}, i.e., no optimization for hardware characterization or debugging for the device with the Fake-Kolkata error map, we have $d=94$, a number of circuit operations of 271, and number of non-local gates of 126.
The transpilation with \textit{optimization\_level=3} for the device with the Fake-Kolkata error map, we have $d=40$ and a number of circuit operations of 108, and number of non-local gates of 24, see the Figure \ref{Circuit-Virtual-n12}-(b). 
In Figure \ref{Circuit-Virtual-n12}-(c), highlighted in red, the transpilation mirrors the ring graph onto the device. Due to the size of the figure, the symbols representing the measurements on each qubit have been arranged in a single column.
Note that by optimizing the output circuit we have significantly reduced the depth and number of circuit operations, and consequently we will mitigate errors during processing. The Figure \ref{Plot-n-12-Depth} shows the distinct levels of optimization for depth. Note that we obtain a good result for the depth with \textit{optimization\_level=1} and that it does not vary for the other levels. This is also valid for the number of circuit operations and the number of non-local gates.
This is because the symmetry of the ring graph perfectly reflects the topology of the device.

\section{Results}
To analyze the noise through the error maps of the fake devices via AerSimulator, we examined the expectation value of F*, the approximation ratio for ring graphs of sizes $n=4,6,8,10$ and $12$, and the probability of success for $n=4$ for the Fake-Washington.
The option for a ring graph in these dimensions is motivated by the possibility of working with a symmetry that perfectly mirrors the topology of the quantum device, in this case for $n=12$ and otherwise for $n<12$. Another essential point to mention is that due to the low connectivity of the ring graph problem, we can explore deeper ansatz circuits, for example, $p=8,9$, and $10$, and observe the noise behavior with increasing circuit depth through the approximation ratio.

The ring graph of size $n=4$ and $6$ was investigated for fake-Lagos with seven qubits. For fake-Kolkata, with 27 qubits, the ring graph of size $n=4, 6, 8, 10$, and $12$ was analyzed. For fake-Washington, with 127 qubits, the ring graph of size $n=4,6,8,10$ and $12$ was analyzed.
In this article, we have used constrained optimization by linear approximation (COBYLA), an optimization method for constrained problems where the objective function's derivatives are unknown \cite{cobyla}. For more details about different classical optimizers  when used with QAOA, see Ref.~\cite{Pendas-2022}.

We have run QAOA 10 times using the default parameters of the Qiskit Aer(AerSimulator) implementation with the density\_matrix method, coupling\_map, noise\_model, and basis\_gates for Fake-Lagos, Fake-Kolkata and Fake-Washington devices. All simulations were with $shots=50,000$ and using the classic COBYLA optimizer, with ring graphs of sizes $n=4,6,8,10$ and $n=12$, with values of $p=\left[ 1,10 \right]$. We computed both the expectation value of F* and the approximation ratio obtained by QAOA. We have achieved the same pattern for the theoretical case (Figure \ref{plot-n2-QAOA}), with the probability of success, using the EM methods on the grid ($\beta,\gamma$) for the Fake-Washington error map and $n=4$.

\subsection{The Fake-Lagos}
We have performed numerical experiments using the Qiskit Aer framework with a local machine simulator (AerSimulator) without and with EM for a ring graph model for $n=4,6$. Note that the Fake-Lagos (Figure \ref{ibm-lagos-error}) has a connectivity architecture for seven qubits. Thus, the QAOA ansatz circuit for the ring graph with $n=4,6$ lacks a complete mapping, requiring additional gates to model the problem.
\begin{figure}[h]
	\centering
	\includegraphics[height=60mm, width=80mm]{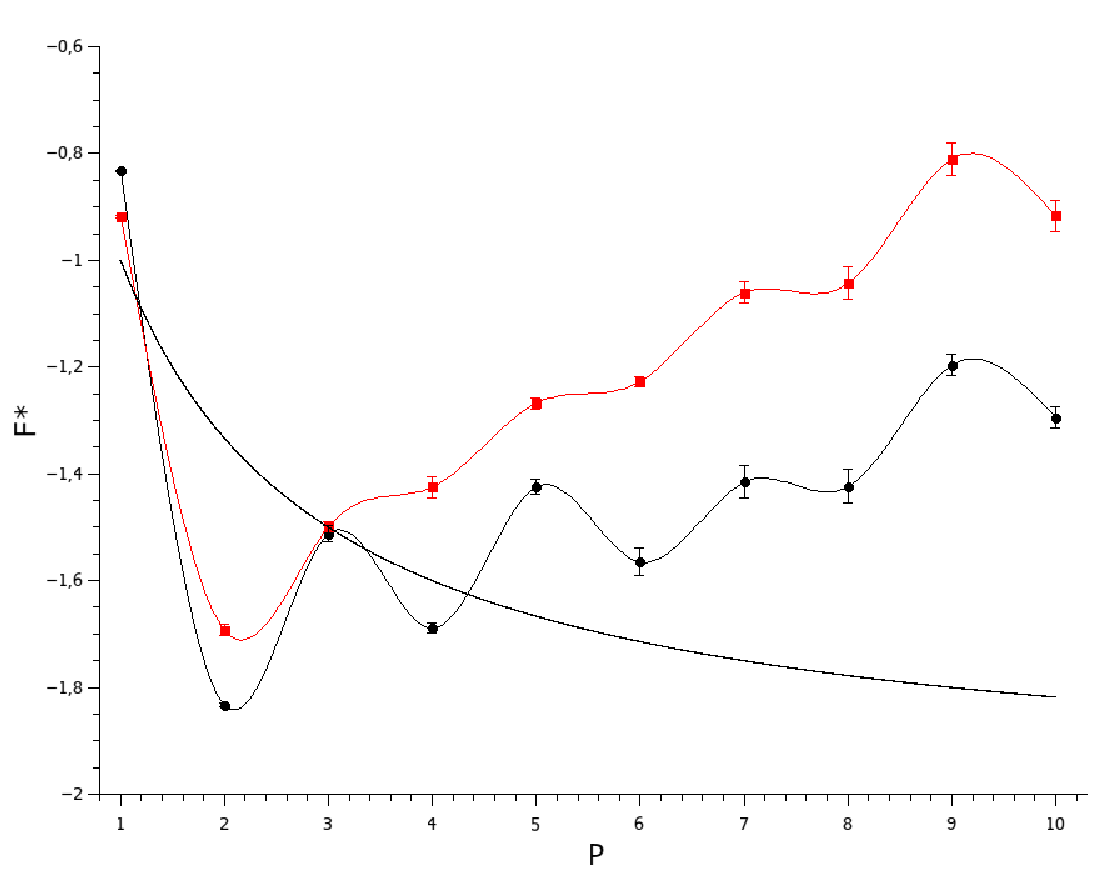}
	\caption{Plot of expectation energy $F^*$ for the ring model $(n=4)$ QAOA states for $P$ layers on AerSimulator with the red (black) points represents simulations without (with) error mitigation. The black line represents the theoretical solution to the problem (see, Eq. \ref{F*-ring}), using the Fake-Lagos error map.}
	\label{Plot-F-StarLagos}
\end{figure}
\begin{figure}[h]
	\centering
	\includegraphics[height=60mm, width=80mm]{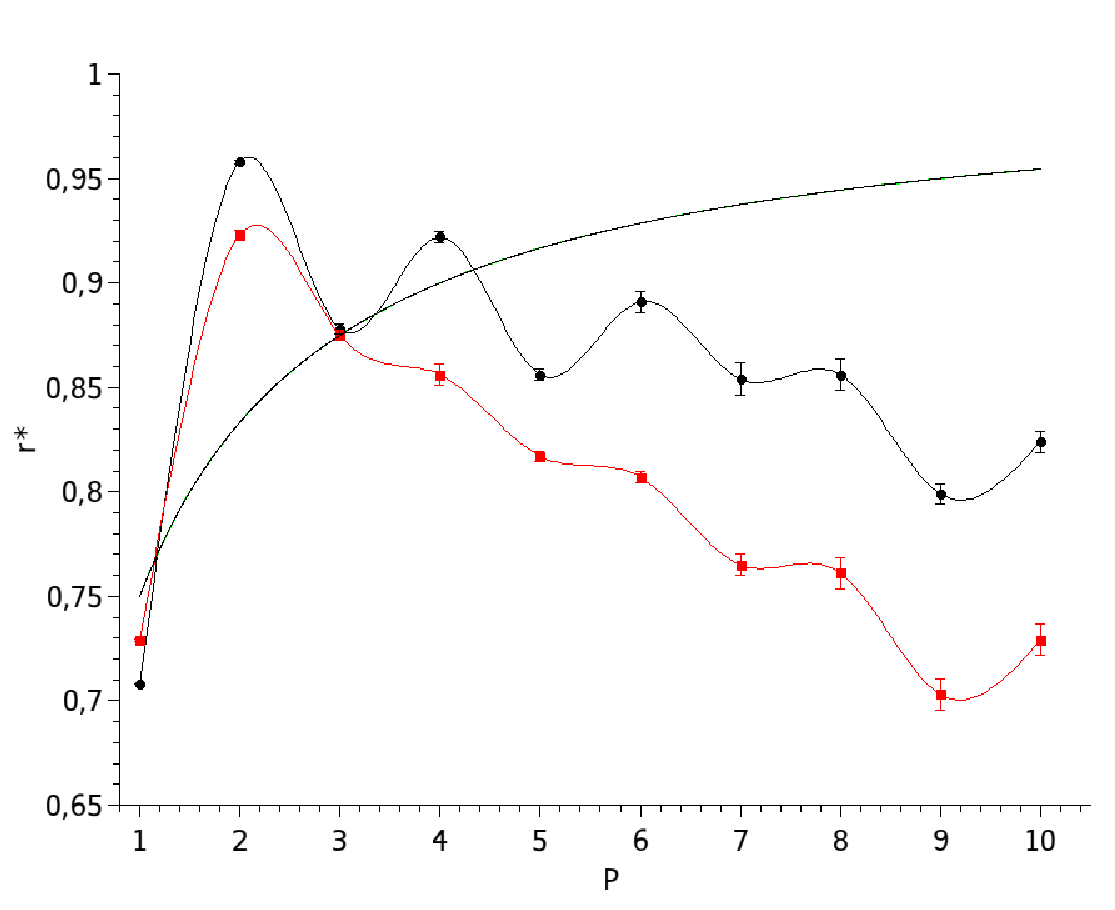}
	\caption{Using the Fake-Lagos error map, we have plotted the approximation ratio for the ring model (n=4) QAOA states for $P$ layers on AerSimulator with the red (black) points representing simulations without (with) error mitigation. The black line represents the theoretical solution to the problem, see Eq.\ref{r*-ring}.}
	\label{Plot-r-StarLagos}
\end{figure}
For this noise model, we calculated the expectation value of F* and the approximation ratio for the ring with ($n=4$) as a function of the number of layers P, Figures \ref{Plot-F-StarLagos} and \ref{Plot-r-StarLagos}, respectively. Note that the noise significantly affects the expectation value of F* and the approximation ratio when we do not use error mitigation methods, Figures \ref{Plot-F-StarLagos} and \ref{Plot-r-StarLagos} (red points lines). 
In this case, the ring graph symmetry mapping can not be applied for $n=4$. Therefore, from Figures \ref{Plot-F-StarLagos} and \ref{Plot-r-StarLagos}, we can observe that even using the error mitigation methods (black points lines), we still have a performance loss with the increase in layers P.
\begin{figure}[h]
	\centering
	\includegraphics[height=60mm, width=80mm]{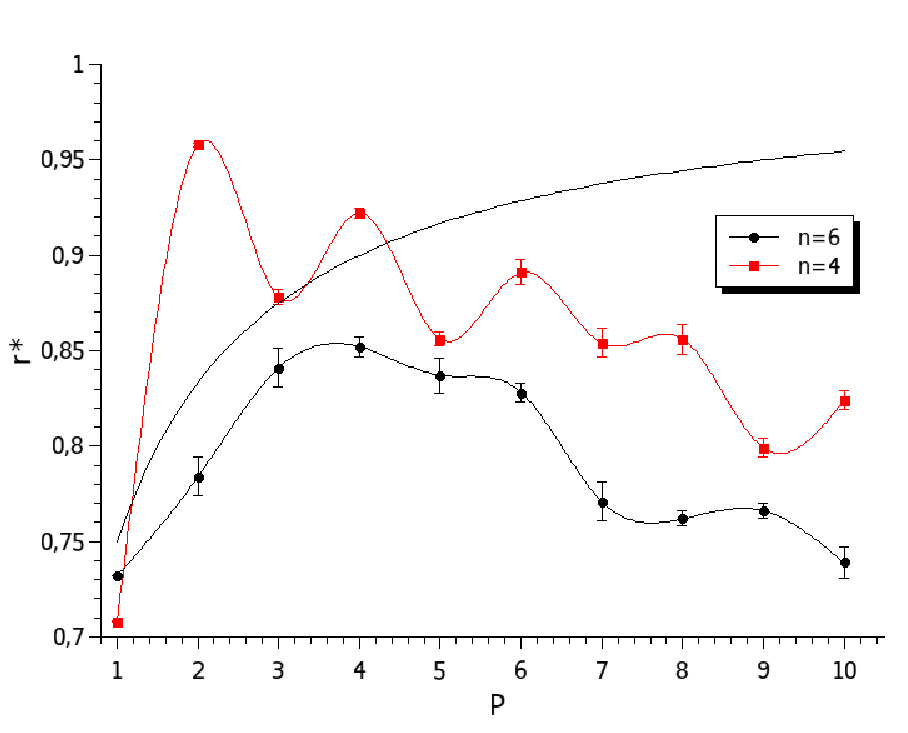}
	\caption{This is a plot of the approximation rate for the QAOA states of the ring graph model, with $n = 4, 6$ for $P$ layers on AerSimulator for Fake-Lagos device. The points represent the error-mitigating simulations for all ring sizes. The black line represents the theoretical solution to the problem, see Eq.\ref{r*-ring}.}
	\label{Plot-r-total-Lagos}
\end{figure}
In Figure \ref{Plot-r-total-Lagos}, we compute the approximation ratio for the $n = 4,6$ ring graphs employing local simulations using the error map of the 7-qubit Fake-Lagos device.
\begin{table}[ht]
	\caption{The data table for the r*-approximation ratio and standard deviation for the ring graphs, n=4, 6, using error mitigation methods and the Fake-Lagos device.}
	{\begin{tabular}{@{}ccccc@{}} \toprule
			P layers &r* $(n=4)$ & Erro & r* $(n=6)$ & Erro \\
			\colrule
			1.0\hphantom{00} & 0.708 & \hphantom{0}0.0003 & 0.732   & \hphantom{00}0.0002       \\
			2.0\hphantom{00} & 0.958 & 0.001              & 0.784   & 0.01        \\
			3.0\hphantom{00} & 0.878 & 0.004              & 0.841   & 0.01       \\
			4.0\hphantom{00} & 0.922 & 0.003              & 0.852   & \hphantom{0}0.005        \\
			5.0\hphantom{00} & 0.856 & 0.004              & 0.837   & \hphantom{0}0.009        \\
			6.0\hphantom{00} & 0.891 & 0.007              & 0.828   & \hphantom{0}0.005       \\
			7.0\hphantom{00} & 0.854 & 0.007              & 0.771   & 0.01        \\
			8.0\hphantom{00} & 0.856 & 0.008              & 0.762   & \hphantom{0}0.004         \\
			9.0\hphantom{00} & 0.799 & 0.005              & 0.766   & \hphantom{0}0.004        \\
			10.0\hphantom{000}& 0.824 & 0.005              & 0.739   & \hphantom{0}0.008        \\ 
			\botrule
		\end{tabular} \label{tab1-fakelagos}}
\end{table} 
In this case, we can not explore the symmetry of the ring graphs with the device topology. For n=6, we have an increase in noise in the QAOA ansatz circuit due to the increase in the size of the ring and the consequent increase in the depth of the circuit. This scenario worsens for n=6, with the approximation ratio decreasing as the number of layers P increases. For example, for P=10, n=6 we have r*=0.739(8) or 73.9\% and for n=4 we have r*=0.824(5) or 82.4\%. See Table \ref{tab1-fakelagos}.

\subsection{The Fake-Kolkata}
We perform numerical experiments using AerSimulator without and with EM for a ring graph model for $n=12$ and using the Fake-Kolkata device (Figure \ref{ibm-lagos-error}) as a noise model. Note that the ring graph can be mirrored in the topology of this device, providing a way to optimize the number of operations in the QAOA circuit.
\begin{figure}[h]
	\centering
	\includegraphics[height=60mm, width=80mm]{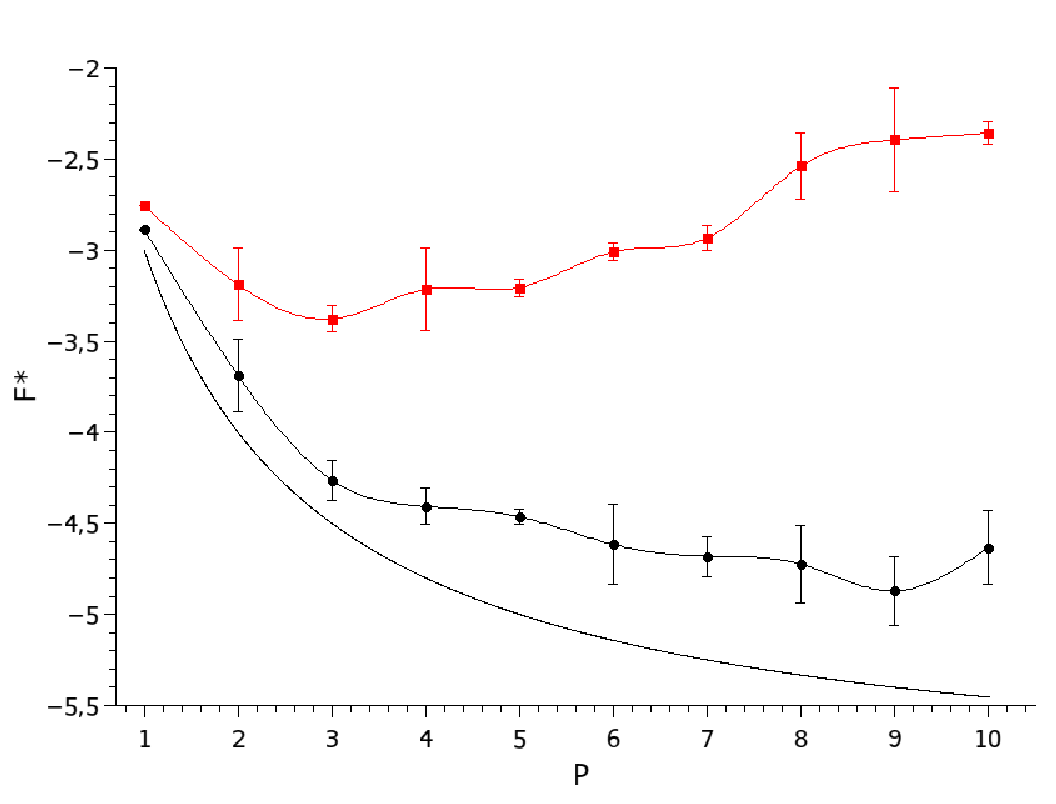}
	\caption{Plot of expectation energy for the ring model $(n=12)$ QAOA states for $P$ layers on AerSimulator with the red (black) points represents simulations without (with) error mitigation. The black line represents the theoretical solution to the problem (see, Eq. \ref{F*-ring}), using the Fake-Kolkata error map.}
	\label{Plot-Kolkata-Simulator}
\end{figure}
\begin{table*}[t]
	\caption{The data table for the r*-approximation ratio and standard deviation for the ring graphs, $n=4,6,8, 10,$ and $12$ using error mitigation methods and the Fake-Kolkata device.}
	{\begin{tabular}{@{}ccccccccccc@{}} \toprule
			P layers & r* $(n=4)$ & Erro & r* $(n=6)$ & Erro & r*$(n=8)$ & Erro  & r*$(n=10)$  & Erro  & r*$(n=12)$ & Erro   \\
			\colrule
			1.0  & 0.7350 &	0.0002 & 0.7230& 0.0003 & 0.7360& 0.0001 & 0.7100& 0.0004 &	0.7406 &0.0002 \\
			2.0  & 0.9490 &	0.0005 & 0.78  & 0.01	& 0.74	& 0.01	 & 0.74  & 0.01	  & 0.80   &0.01   \\
			3.0	 & 0.915  &	0.002  & 0.719 & 0.005	& 0.80	& 0.01	 & 0.82  & 0.01	  & 0.855  &0.009   \\
			4.0	 & 0.898  &	0.002  & 0.845 & 0.008	& 0.815	& 0.008	 & 0.850 & 0.006  & 0.867  &0.008  \\
			5.0	 & 0.868  &	0.002  & 0.794 & 0.005	& 0.804	& 0.009	 & 0.831 & 0.003  &	0.872  &0.003   \\
			6.0	 & 0.857  &	0.005  & 0.803 & 0.003	& 0.76	& 0.01	 & 0.84  & 0.01	  & 0.88   &0.01  \\
			7.0	 & 0.832  &	0.005  & 0.777 & 0.006	& 0.759	& 0.007	 & 0.833 & 0.005  &	0.890  &0.009  \\
			8.0	 & 0.808  &	0.007  & 0.767 & 0.006	& 0.75 	& 0.01	 & 0.885 & 0.009  &	0.893  &0.01   \\
			9.0  & 0.791  &	0.001  & 0.72  & 0.01	& 0.718	& 0.007	 & 0.88  & 0.02	  & 0.90   &0.01   \\ 
			10.0 & 0.76   &	0.01   & 0.714 & 0.005	& 0.713	& 0.007	 & 0.874 & 0.009  &	0.88   &0.01  \\
			\botrule
		\end{tabular} \label{tab2-fakekolkata}}
\end{table*}
\begin{figure}[h]
	\centering
	\includegraphics[height=60mm, width=80mm]{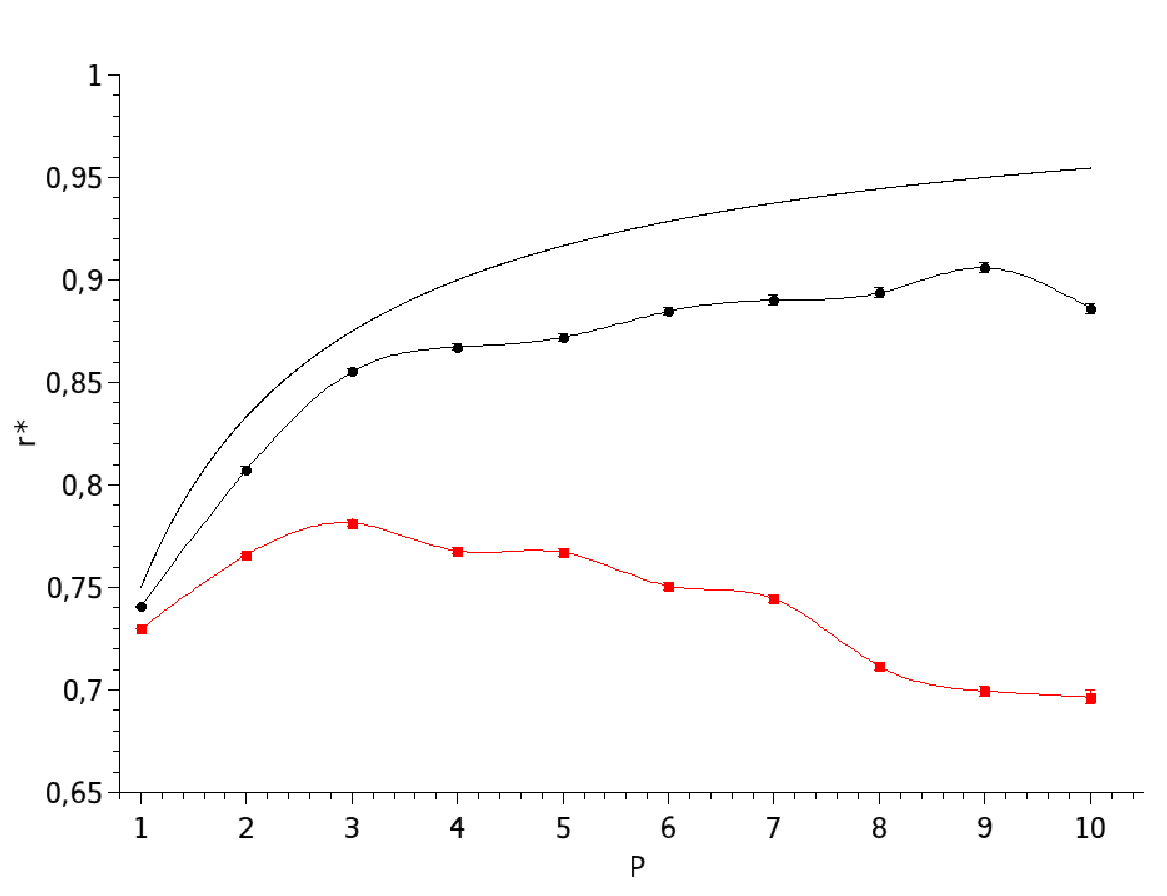}
	\caption{Using the Fake-Kolkata error map, we have plotted the approximation ratio for the ring model $(n=12)$ QAOA states for $P$ layers on AerSimulator with the red (black) points representing simulations without (with) error mitigation. The black line represents the theoretical solution to the problem, see Eq.\ref{r*-ring}.}
	\label{Plot-r-Kolkata}
\end{figure}

We calculate the expectation value of F* and the r*-approximation ratio for the ring with $n=12$ as a function of the number of layers P, Figures \ref{Plot-Kolkata-Simulator} and \ref{Plot-r-Kolkata} , respectively. Note that noise also significantly affects the expectation of F* and r* when we do not use error mitigation methods, Figures \ref{Plot-Kolkata-Simulator} and \ref{Plot-r-Kolkata}, red dots lines. However, we now have the symmetry of the ring graph reflected in the device topology. Therefore, we can observe that using the error mitigation methods, we experience an increase in performance regarding P for F* and r*, Figures \ref{Plot-Kolkata-Simulator} and \ref{Plot-r-Kolkata}, black dots lines, compared to the solution of Eq. \ref{r*-ring}, black line.

\begin{figure}[h]
	\centering
	\includegraphics[height=60mm, width=80mm]{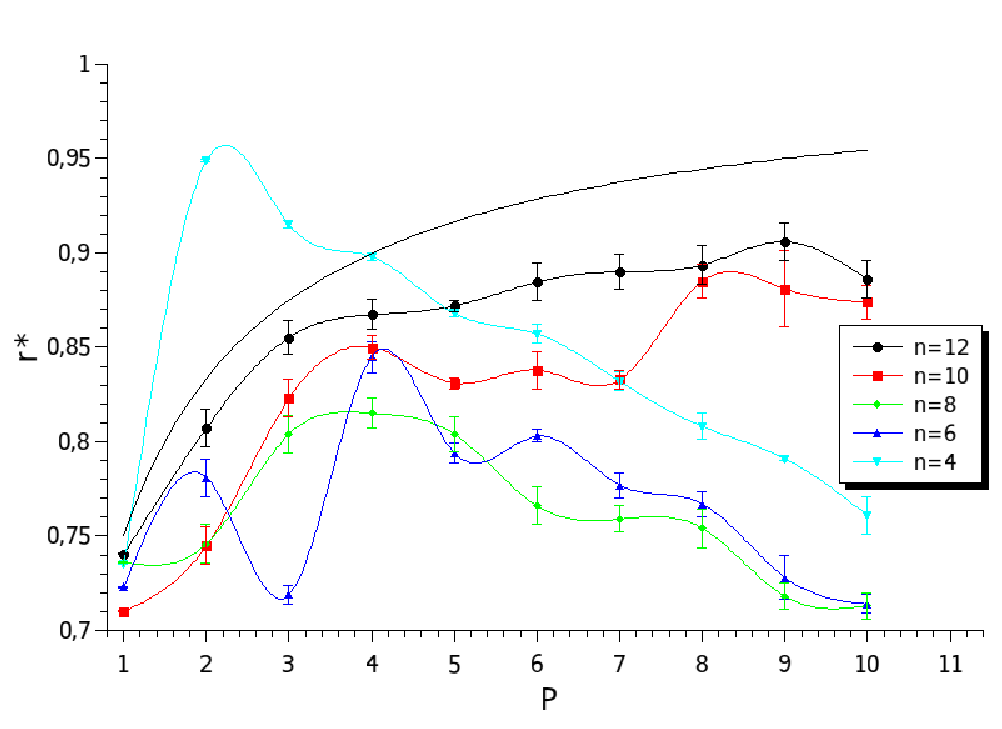}
	\caption{This is a plot of the approximation rate for the QAOA states of the ring graph model, with $n = 4, 6, 8, 10, 12$ for $P$ layers on AerSimulator for Fake-Kolkata device. The points represent the error-mitigating simulations for all ring sizes. The black line represents the theoretical solution to the problem, see Eq.\ref{r*-ring}.}
	\label{Plot-r-total-FakeKolkata}
\end{figure}

In Figure \ref{Plot-r-total-FakeKolkata}, we compute the approximation ratio for the $n = 4, 6, 8, 10,$ and $12$ ring graphs employing local simulations using the error map of the 27-qubit Fake-Kolkata device.
In this case, we can explore the symmetry of the ring graph $(n=12)$ with the device topology. As shown in Figure \ref{Plot-r-total-FakeKolkata}, the best result for the approximation ratio is obtained for the ring graph $n=12$. However, for $n=10$, we have results for the approximation ratio close to the values of the ring graph n=12 for layers $P=8, 9,$ and $10$. For the ring graphs n=4, 6, and 8, we obtain similar results with a particular emphasis on n=4, where we receive a better outcome for layers $P=8, 9,$ and $10$. See Table \ref{tab2-fakekolkata} for more details.

\subsection{The Fake-Washington}
For Fake-Washington (Figure \ref{ibm-washington-error}), we performed numerical experiments using AerSimulator without and with EM for a ring graph model for $n=4, 6, 8, 10$ and $12$. 
Note that for $n=12$, the symmetry of the ring graph can be perfectly reflected in the topology of this device. Therefore, the QAOA ansatz circuit for this graph can achieve the best optimization concerning the number of operations.
\begin{figure}[h]
	\centering
	\includegraphics[height=60mm, width=80mm]{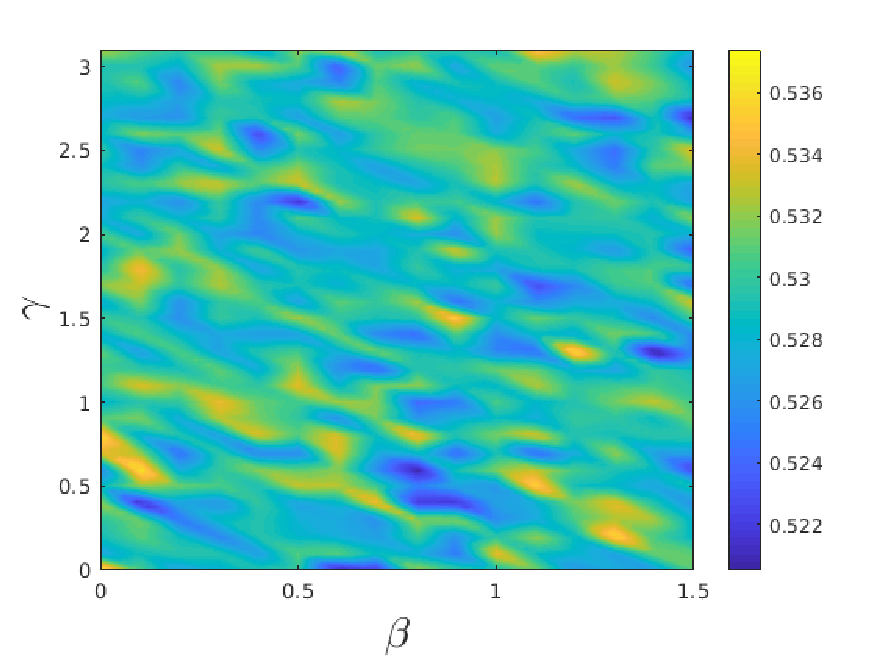}
	\caption{Contour plot of the probability of success for the ring model QAOA states for n=4 and $p=1$. These results were obtained with AerSimulator, without noise mitigation, using the Fake-Washington error map.}
	\label{Plot-Prob-Noise}
\end{figure}
\begin{figure}
	\centering
	\includegraphics[height=60mm, width=80mm]{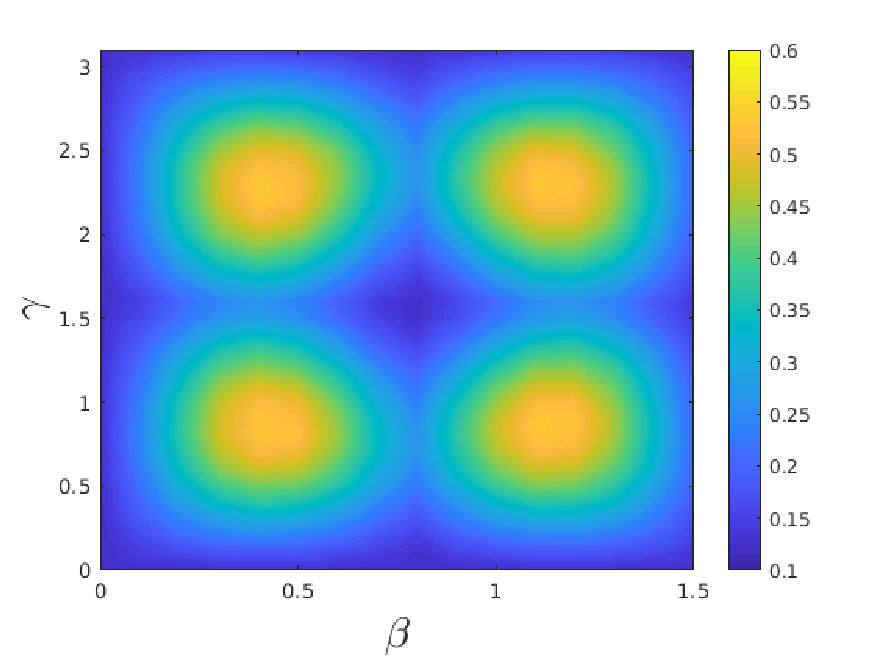}
	\caption{Contour plot of the probability of success for the ring model QAOA states for n=4 and $p=1$. These results were obtained with AerSimulator, with error mitigation, using the Fake-Washington error map.}
	\label{Plot-Prob-EM}
\end{figure}

\begin{figure}
	\centering
	\includegraphics[height=60mm, width=80mm]{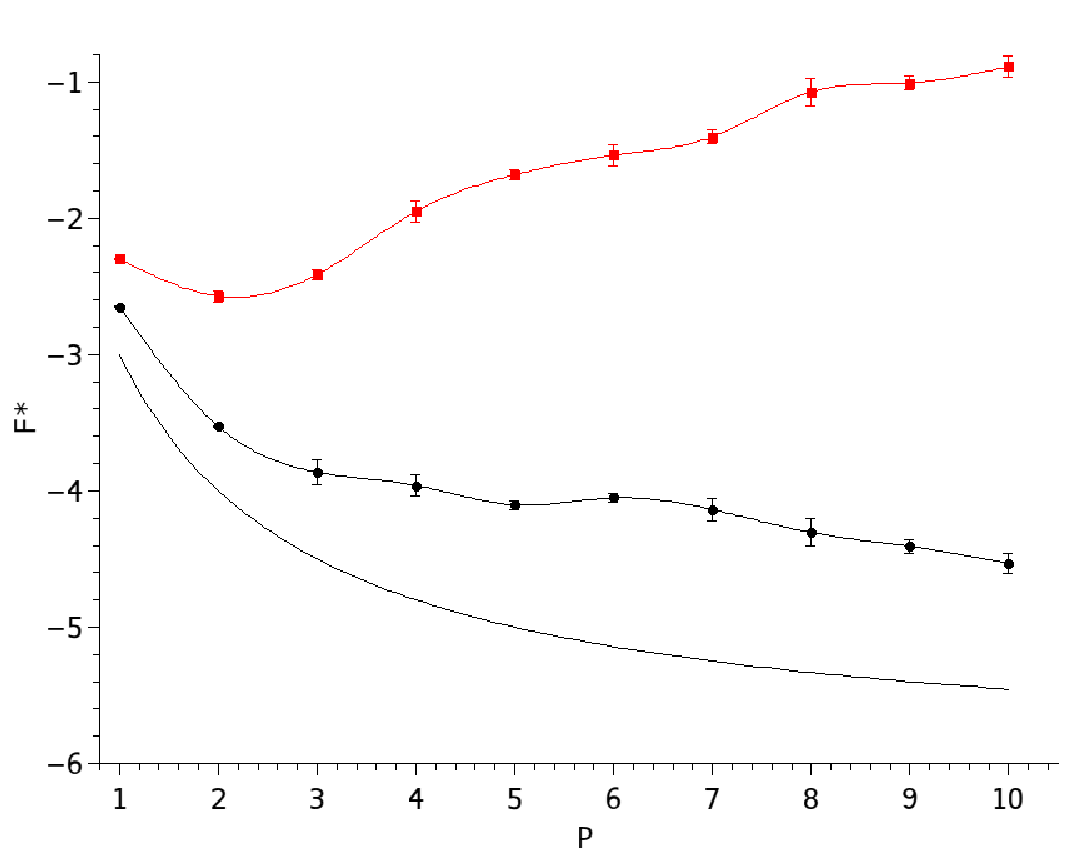}
	\caption{Plot of expectation energy for the ring model (n=12) QAOA states for $P$ layers on AerSimulator with the red (black) points represents simulations without (with) error mitigation. The black line represents the theoretical solution to the problem (see, Eq. \ref{F*-ring}), using the Fake-Washington error map.}
	\label{Plot-Washington-Simulator}
\end{figure}

Figures \ref{Plot-Prob-Noise} and \ref{Plot-Prob-EM} show the probability of success (probability of finding the ground state), after applying QAOA for p=1 as a function of $\gamma$ and $\beta$ for the max-cut of the ring model with n=4. We performed 50.000 shots with AerSimulator, without and with noise mitigation, respectively, using the Fake-Washington error map, with the angles $\beta_k$ and $\gamma_k$ varying by $[0,\pi/2]$ and $[0,\pi]$, respectively. We ran a grid search on the AerSimulator for $p=1$ with resolution $\pi/30$.
\begin{figure}[h]
	\centering
	\includegraphics[height=60mm, width=80mm]{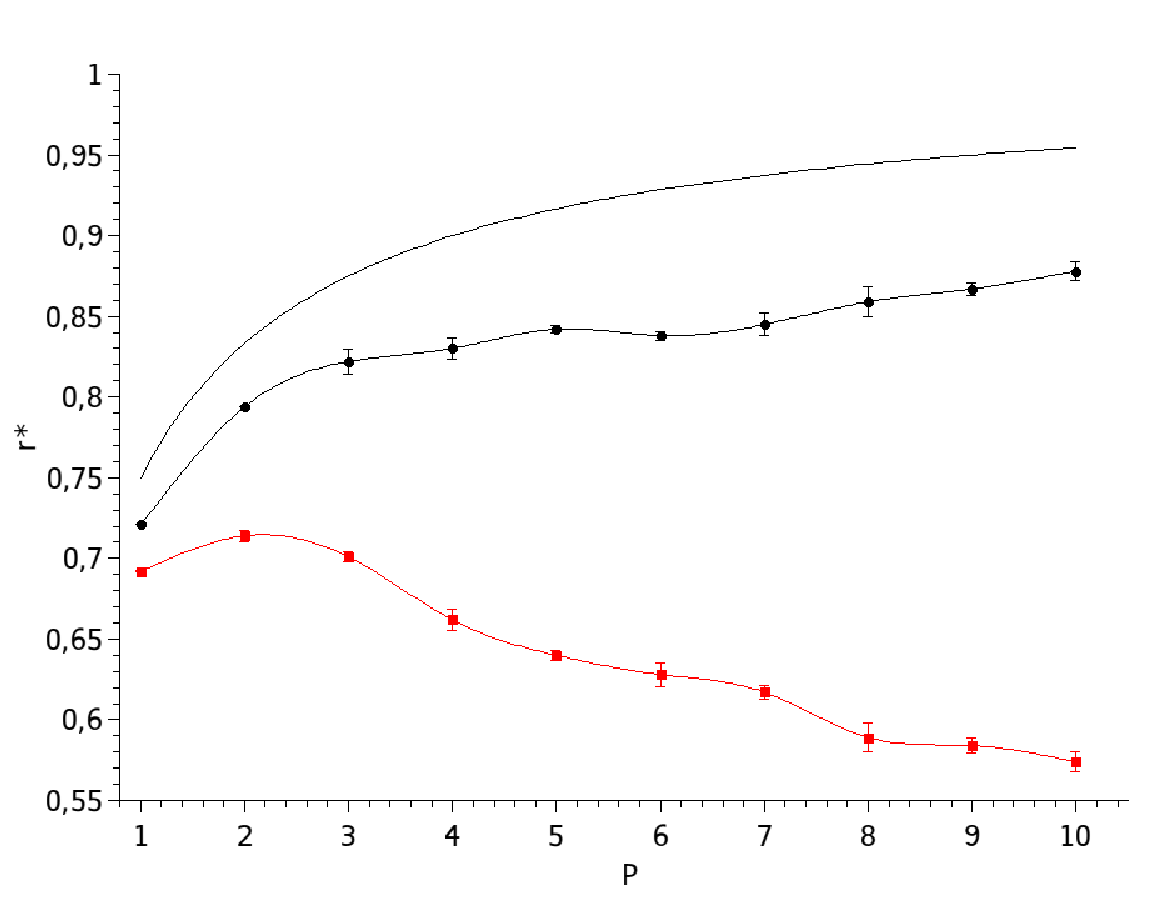}
	\caption{Using the Fake-Washington error map, we have plotted the approximation ratio for the ring model (n=12) QAOA states for $P$ layers on AerSimulator with the red (black) points representing simulations without (with) error mitigation. The black line represents the theoretical solution to the problem,see Eq.\ref{r*-ring}.}
	\label{Plot-r-Washington}
\end{figure}
\begin{figure}[h]
	\centering
	\includegraphics[height=60mm, width=80mm]{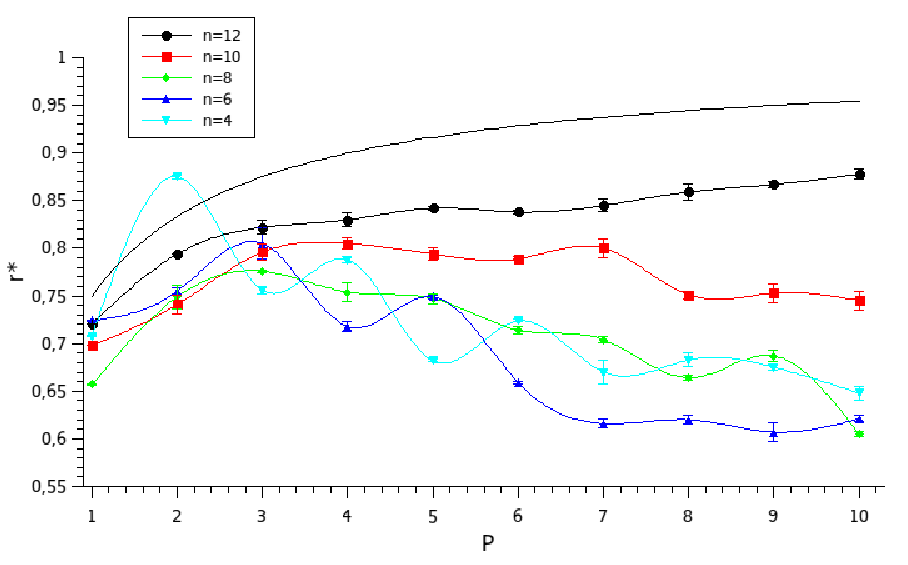}
	\caption{This is a plot of the approximation rate for the QAOA states of the ring model, with $n = 4, 6, 8, 10, 12$ for $P$ layers on AerSimulator. The points represent the error-mitigating simulations for all ring sizes. The black line represents the theoretical solution to the problem, see Eq.\ref{r*-ring}.}
	\label{Plot-r-total-Washington}
\end{figure}
In addition, we observe that due to noise, Figure \ref{Plot-Prob-Noise} does not show any regions with an increase in the probability of success for the ($\gamma,\beta$) grid or the expectation energy value. Using error mitigation techniques, Figure \ref{Plot-Prob-EM} shows an entirely different scenario for the probability of success. We can identify regions of high probability with respect to the ($\gamma^*$,$\beta^*$) values, in particular recovering analytically obtained results through Eq. \ref{n-2-Fp-f} and Figure \ref{plot-n2-QAOA}.
\begin{table*}[t]
	\caption{The data table for the r*-approximation ratio and standard deviation for the ring graphs, $n=4,6,8, 10,$ and $12$ using error mitigation methods and the Fake-Washington device.}
	{\begin{tabular}{@{}ccccccccccc@{}} \toprule
			P layers & r* $(n=4)$ & Erro & r* $(n=6)$ & Erro & r*$(n=8)$ & Erro  & r*$(n=10)$  & Erro  & r*$(n=12)$ & Erro   \\
			\colrule
			1.0  &0.7080 &	0.0002 &   0.7240 &	0.0003 & 0.6570 & 0,0001 & 0.6980&	0.0001 & 0.72100 & 0.00008     \\
			2.0	 & 0.875 &	0.002  &	0.755 &	0.003  & 0.74   & 0.01	 & 0.741 &	0.01   & 0.7940  & 0.0001     \\
			3.0	 & 0.756 &	0.003  &	0.80  &	0.01   & 0.776  & 0.001  & 0.796 &	0.008  & 0.822   & 0.007     \\
			4.0	 & 0.788 &	0.003  &	0.718 &	0.005  & 0.75   & 0.01	 & 0.805 &	0.006  & 0.830   & 0.006     \\
			5.0  & 0.682 &	0.001  &	0.749 &	0.001  & 0.747  & 0.005	 & 0.794 &	0.007  & 0.842   & 0.002      \\
			6.0  & 0.724 &	0.001  &	0.659 &	0.001  & 0.714  & 0.003  & 0.789 &	0.002  & 0.838   & 0.002     \\
			7.0	 & 0.67  &	0.01   &	0.616 &	0.005  & 0.704  & 0.003  & 0.80  &  0.01   & 0.845   & 0.006     \\
			8.0	 & 0.683 &	0.007  &	0.620 &	0.005  & 0.664  & 0.002	 & 0.751 &	0.004  & 0.859   & 0.008     \\
			9.0	 & 0.675 &	0.003  &	0.60  &	0.01   & 0.687  & 0.006  & 0.753 &	0.01   & 0.867   & 0.003     \\
			10.0 & 0.648 & 	0.007  &    0.621 &	0.003  & 0.605  & 0.002	 & 0.745 &	0.01   & 0.878   & 0.005     \\
			\botrule
		\end{tabular} \label{tab2-fakewashington}}
\end{table*} 
We calculate the expectation value of F* and the r* approximation ratio for the ring with $n=12$ as a function of the number of layers P, Figures \ref{Plot-Washington-Simulator} and \ref{Plot-r-Washington}, respectively. Note that noise also significantly affects the expectation of F* and r* when we do not use error mitigation methods, Figures \ref{Plot-Washington-Simulator} and \ref{Plot-r-Washington}, red lines. However, we now have the symmetry of the ring graph reflected in the device topology. Therefore, we can observe that using the error mitigation methods, we experience an increase in performance regarding P for F* and r*, Figures \ref{Plot-Washington-Simulator} and \ref{Plot-r-Washington}, black dots lines, compared to the solution of Eq. \ref{r*-ring}, black line curve.

In Figure \ref{Plot-r-total-Washington}, we calculate the approximation ratio for QAOA states for the following ring graphs of sizes $n=4, 6, 8, 10,$ and $12$ as a function of $P$ layers. The points represent the error-mitigating simulations for all ring sizes.
The conclusion from this figure is that as the depth of the circuit increases, i.e., as $P$ increases, we have an increase in the number of operations in the circuit, which results in more noise. The approximation ratio is reduced despite the use of error mitigation methods. Contrary to this result, we have the approximation ratio for the ring graph $n=12$, which performs better than the other curves in this figure. This is due to the symmetry of the ring, which reflects the topology of the device and thus plays a fundamental role in approximating the theoretical curve. See Table \ref{tab2-fakewashington} for more details.

\section{Conclusion}
In this paper, we conduct a study of the error maps of real quantum devices and their influence on the performance of the quantum approximate optimization algorithm. We perform experiments on the max-cut problem for different sizes of ring graphs using a local simulator (AerSimulator), with and without error mitigation through fake devices. Using QAOA for the max-cut problem, we investigate the influence of the symmetry of the ring graphs in determining the approximation ratios, the expectation value of the energy, and the probability of success and compare them with theoretical results from the literature. Our results demonstrated that error mitigation methods based on the Qiskit framework are essential for better results in the presence of noise. It is important to highlight that when the symmetry of the ring mirrors the topology of the device, there is a significant improvement in the results due to a reduction in the number of circuit operations.

Our work has also presented an alternative approach for developers who need access to relatively expensive real quantum devices. This approach allows developers to build, test, and debug their code locally without relying on cloud resources. In this initiative, we use fake devices that are error maps (noise models) of the following devices: fake-Lagos, fake-Kolkata, and fake-Washington, with 7, 27, and 127 qubits, respectively.

On the other hand, we show that even for ring graphs, we notice quite different behavior between the analytical approximation ratio formula and the results obtained by the quantum approximate optimization algorithm. This opens, then, two interesting avenues for future investigation. First, the possibility of using other QAOA ansatz to reduce noise and thus guide the simulations in the search for better results. Second, the use of simplifications and equivalences of quantum circuits to reduce the number of circuit operations when possible, especially in the absence of symmetry of graphs with the topology of the quantum device.


\end{document}